\documentclass[runningheads]{llncs}
\usepackage[T1]{fontenc}

\usepackage{graphicx}

\usepackage{times}
\usepackage{soul}
\usepackage{url}
\usepackage[utf8]{inputenc}
\usepackage{amsmath}
\usepackage{amsthm}
\usepackage{booktabs}
\usepackage{algorithm}
\usepackage{algpseudocode}

\usepackage{makecell}

\usepackage[switch]{lineno}

\usepackage{graphicx}
\usepackage{parskip}
\usepackage{graphicx}
\usepackage{comment}
\usepackage{hyperref}
\usepackage{subcaption}
\usepackage{xcolor}

\usepackage{booktabs}
\usepackage{multirow}

\usepackage{nicefrac}

\usepackage{algorithm}

\usepackage{hyperref}
\usepackage{comment}

\begin{document}

\title{Expanding Chemical Representation with k-mers and Fragment-based Fingerprints for Molecular Fingerprinting}
%
%
\author{Sarwan Ali
\and
Prakash Chourasia
\and
Murray Patterson
\\
\{sali85, pchourasia1\}@student.gsu.edu, mpatterson30@gsu.edu
}
\authorrunning{S. Ali et al.}
%
\institute{Georgia State University, Atlanta GA, 30303, USA 
}
\maketitle              

\begin{abstract}

This study introduces a novel approach, combining substruct counting, $k$-mers, and Daylight-like fingerprints, to expand the representation of chemical structures in SMILES strings. The integrated method generates comprehensive molecular embeddings that enhance discriminative power and information content. Experimental evaluations demonstrate its superiority over traditional Morgan fingerprinting, MACCS, and Daylight fingerprint alone, improving chemoinformatics tasks such as drug classification. The proposed method offers a more informative representation of chemical structures, advancing molecular similarity analysis and facilitating applications in molecular design and drug discovery. It presents a promising avenue for molecular structure analysis and design, with significant potential for practical implementation.


\end{abstract}

\section{Introduction}
\begin{figure}[h!]
  \centering
  \includegraphics[scale=0.35]{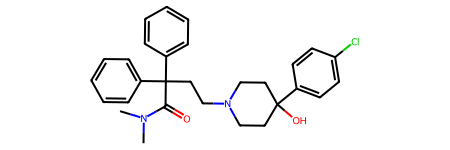}
  \caption{Molecular structure for the drug named ``Loperamide", with
    solubility AlogPS (Aqueous solubility and Octanol/Water partition coefficient) value of 0.00086, and the following SMILES
    string: {\hspace*{.5cm}
      \texttt{CN(C)C(=O)C(CCN1CCC(O)(CC1)C1=CC=C\\(Cl)C=C1)(C1=CC=CC=C1)C1=CC=CC=C1}}}
  \label{fig_molecule_1}
\end{figure}

Molecular structure analysis is a vital endeavor in drug discovery and molecular design~\cite{sellwood2018artificial}. Due to their simplicity and usability, Simplified Molecular Input Line Entry System (SMILES) strings have become more popular as a preferred way for encoding molecular structure data~\cite{schwaller2022machine} (see Figure~\ref{fig_molecule_1} for an example of a SMILES string). However, modeling and analyzing molecular structures expressed as SMILES strings present several difficulties~\cite{krenn2020self}. These difficulties include managing the enormous complexity of the data and comprehending the intricate non-linear interactions between the structures. Applications in machine learning rely primarily on numerical representations of the data~\cite{Glorot2011domain}.
The conversion of SMILES strings into machine-readable numerical representations is a complex task that demands sophisticated techniques.

The analysis of SMILES strings has gained significant importance in the field of drug discovery and cheminformatics~\cite{chen2018cheminformatics}. SMILES strings are a well-liked method for encoding molecular information in machine learning models because they offer a succinct description of a molecule's structure~\cite{wigh2022review,ucak2023reconstruction}. These models are used for several tasks, including subtype prediction~\cite{choi2020target} and drug solubility prediction~\cite{francoeur2021soltrannet}. 
By comparing the effectiveness of various embedding techniques and ML models for classification tasks using SMILES strings as input, this research intends to close this knowledge gap. The project also suggests a fresh approach to SMILES string analysis. The results of this study may have important ramifications for drug discovery and aid in determining the best techniques for predicting molecular characteristics.


The proposed approach addresses challenges in modeling and analyzing chemical structures represented as SMILES strings. It incorporates various fingerprinting methodologies to capture intricate non-linear interactions and overcome high-dimensional data. Using the RDKit library, we transform SMILES strings into molecular structures and generate feature vectors. To gather more information, we combine the Morgan fingerprint with $k$-mers extracted from the SMILES string. Which helps to capture local and variable-length substructs, revealing structural relationships and functional groups. The effectiveness of the proposed fingerprint embeddings is evaluated in drug subcategory prediction tasks.

The proposed method has a wide variety of potential applications, including drug discovery, and molecular design. It offers the opportunity to quickly search through vast datasets of chemical structures in search of compounds with desirable properties. By creating low-dimensional embeddings and using them to find molecules with related qualities, the approach makes it possible to construct unique compounds with certain properties. Overall, this signifies a promising avenue for molecular structure analysis, employing kernel methods to unlock new possibilities.
Following are our contributions:
\begin{enumerate}
    \item We propose a novel method for embedding generation for SMILES strings, which can be used for underlying supervised analysis such as classification. Our approach is predicated on the notion of first turning SMILES strings into molecular graphs, and computing fingerprints while incorporating $k$-mers.
    \item We show that the proposed method preserves both the structural and contextual information better when compared to the baselines.
    \item Using extensive experimentation, we demonstrated that the proposed method can achieve higher predictive performance on the benchmark SMILES string dataset.
\end{enumerate}

The remainder of the paper is structured as: Section~\ref{sec_related_work} reviews related work, Section~\ref{sec_proposed_approach} presents our proposed approach, Section~\ref{sec_experimental_setup} describes the dataset and experimental setup, Section~\ref{sec_results} presents the outcomes of the proposed and baseline methods, and Section~\ref{sec_conclusion} concludes the paper.

\section{Related Work}~\label{sec_related_work}

Molecular fingerprints are popular and widely used for encoding structural information in molecules~\cite{probst2018probabilistic,wigh2022review,ucak2023reconstruction}. They have been successfully applied in drug solubility prediction~\cite{nakajima2021machine}, with random forest regression and support vector regression showing superior performance~\cite{chen2018rise}. Graph convolutional neural networks have also achieved promising results~\cite{rupp2012fast,zhang2020automatic}. Further research is needed to explore different embeddings, classification, and regression models for solubility and drug subtype prediction.
Kernel methods, such as kernel ridge regression (KRR)\cite{fabregat2022metric,stuke2019chemical} and support vector machine (SVM)\cite{tkachev2019floating,thomas2017multi}, are commonly used for molecular data analysis. 
To find similarities using molecular fingerprints, several works propose to combine various methods using data fusion~\cite{salim2003combination}, either by combining different fingerprints ~\cite{willett2013fusing,sastry2013boosting,awale2014multi} or by combining fingerprints with other methods, especially structure-based methods~\cite{muegge2016overview}. 
The several combinations help to capture various chemical information, making them more relevant and making it better compared to what a single approach would introduce. 
Kernel principal component analysis (PCA) effectively reduces dimensionality and feature extraction~\cite{rensi2017flexible,fu2011combination}. It has been successfully used in molecular property prediction and activity classification~\cite{fu2011combination}. However, these methods have limitations, such as computational complexity and potential overfitting, especially for large datasets.

\section{Proposed Approach}~\label{sec_proposed_approach}
In this section, we discuss the main idea of the Morgan Fingerprint followed by the integration of $k$-mers in the Morgan Fingerprint.

The Morgan Fingerprint algorithm~\cite{nakajima2021machine}, as shown in Algorithm~\ref{algo_morgan}, is designed to generate a fingerprint representation for a given SMILES string. The fingerprint captures the occurrence of substructs within the SMILES string, which is defined by a specified radius. The algorithm starts by initializing an empty dictionary, substructCnt, to store the occurrence count of each substruct. It then iterates over the specified radius, and for each radius, iterates over the SMILES string to extract substructs of that radius. If a substruct is already present in substructCnt, its count is incremented; otherwise, it is added to substructCnt with an initial count of 1. Once all substructs have been counted, they are sorted alphabetically to create the list sortedSubstruct.
Next, the algorithm constructs the binary fingerprint representation. It initializes an empty list, fingerprint, and iterates over the sorted substructs. For each substruct, its occurrence count is converted into a binary representation using 32 bits, where each bit corresponds to whether the count has a value of 0 or 1. These binary representations are appended to fingerprint.
After constructing the fingerprint, the algorithm checks if the length of the fingerprint is greater than or equal to the desired number of bits, nBits. If it is, the fingerprint is truncated to the first nBits elements. Otherwise, it is padded with zeros ([0]) to reach the desired length.
Finally, the generated fingerprint is returned as the output of the GenerateMorganFingerprint function.
Figure~\ref{Fig_Morgan_Fingerprint} shows the process we use for generating Morgan fingerprints.

\begin{algorithm}[h!]
    \caption{Morgan Fingerprint}
    \label{algo_morgan}
    \begin{algorithmic}[1]
    \scriptsize
        \Function{GenerateMorganFingerprint}{smiles, radius=2, nBits=2048}
            \State substructCnt $\gets$ []
            \For{$i \gets 1$ to radius}
                \For{$j \gets 0$ to len(smiles)-i}
                    \State substruct $\gets$ smiles[j:j+i]
                    \If{substruct $\in$ substructCnt}
                        \State substructCnt[substruct] $\mathrel{+}$ = 1
                    \Else
                        \State substructCnt[substruct] $\gets$ 1
                    \EndIf
                \EndFor
            \EndFor
            \State sortedSubstruct $\gets$ sort(substructCnt.keys())
            \State fingerprint $\gets$ []
            \For{substruct $\in$ sortedSubstruct}
                \State substructBinary $\gets$ [int(bit)  for bit in bin(substructCnt[substruct])[2:].zfill(32)]
                \State fingerprint.extend(substructBinary)
            \EndFor
            \If{len(fingerprint) $\geq$ nBits}
                \State fingerprint $\gets$ fingerprint[:nBits]
            \Else
             \State fingerprint $\gets$ fingerprint + [0] $\times$ (nBits - len(fingerprint))
            \EndIf
            \State \textbf{return} fingerprint
        \EndFunction
    \end{algorithmic}
\end{algorithm}


\subsection{Integration of $k$-mers in Morgan Fingerprint}
The "Morgan Fingerprint with k-mers" algorithm, as depicted in Algorithm~\ref{algo_morgan_kmers}, generates a fingerprint representation for a given SMILES string. The function GenerateMorganFingerprintKmers takes the SMILES string as input along with optional parameters such as the radius (default value of 2), $k$-mer length (default value of 3), and desired number of bits for the fingerprint (default value of 2048).
The algorithm starts by initializing an empty list, substructure count (substructCnt), to store the counts of substructs. It then iterates through each possible radius value from 1 to the specified radius. Within this loop, it further iterates through the characters of the SMILES string to extract substructs of the given radius. The substruct is checked for existence in substructCnt, and if present, its count is incremented; otherwise, a new entry is added with an initial count of 1.
Next, another loop is executed to generate k-mers from the SMILES string. Similar to the previous loop, it extracts substructs of length k from the string and updates their counts in substructCnt.
The algorithm then sorts the substructs in substructure alphabetically to ensure consistent ordering. It initializes an empty list, fingerprint, to store the binary representation of the substruct counts. For each substruct in the sorted order, it converts the corresponding count to a binary representation of length 32 and appends each bit to the fingerprint.
After generating the fingerprint, the algorithm checks if the length of the fingerprint is greater than or equal to the desired number of bits. If it exceeds, the fingerprint is truncated to the desired length; otherwise, it is padded with additional zeros to match the desired length.
Finally, the algorithm returns the generated fingerprint as the output of the function.

\begin{algorithm}[h!]
\caption{Morgan Fingerprint with k-mers}
\label{algo_morgan_kmers}
\begin{algorithmic}[1]
 \scriptsize
\Function{GenerateMorganFingerprintKmers}{smiles, radius=2, k=3, nBits=2048}
\State substructCnt $\gets$ []
\For{$i \gets 1$ to radius}
\For{$j \gets 0$ to len(smiles) - i}
\State substruct $\gets$ smiles[j:j+i]
\If{substruct $\in$ substructCnt}
\State substructCnt[substruct] $\mathrel{+}$= 1
\Else
\State substructCnt[substruct] $\gets$ 1
\EndIf
\EndFor
\EndFor
\For{$j \gets 0$ to len(smiles) - k}
\State substruct $\gets$ smiles[j:j+k]
\If{substruct $\in$ substructCnt}
\State substructCnt[substruct] $\mathrel{+}$= 1
\Else
\State substructCnt[substruct] $\gets$ 1
\EndIf
\EndFor
\State sortedSubstruct $\gets$ sort(substructCnt.keys())
\State fingerprint $\gets$ []
\For{substruct $\in$ sortedSubstruct}
\State substructBinary $\gets$ [int(bit) for bit in bin(substructCnt[substruct])[2:].zfill(32)]
\State fingerprint.extend(substructBinary)
\EndFor
\If{len(fingerprint) $\geq$ nBits}
\State fingerprint $\gets$ fingerprint[:nBits]
\Else
\State fingerprint $\gets$ fingerprint + [0] $\times$ (nBits - len(fingerprint))
\EndIf
\State \textbf{return} fingerprint
\EndFunction
\end{algorithmic}
\end{algorithm}

\subsection{Daylight Fingerprint}
\begin{figure}[h!]
  \centering
  \begin{subfigure}{.33\textwidth}
  \centering
   \includegraphics[scale=0.19]{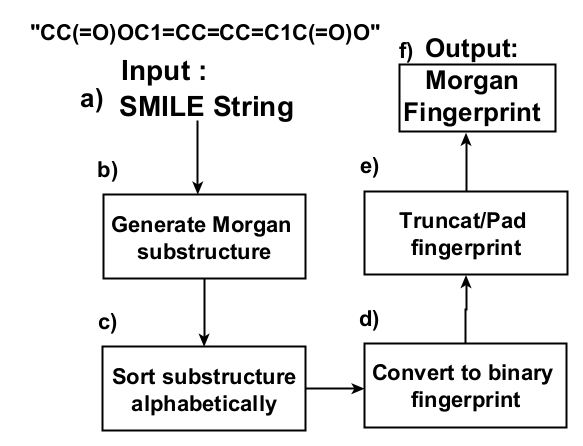}
  \caption{Morgan Fingerprint.}
  \label{Fig_Morgan_Fingerprint}
  \end{subfigure}%
  \begin{subfigure}{.33\textwidth}
  \centering
  \includegraphics[scale=0.19]{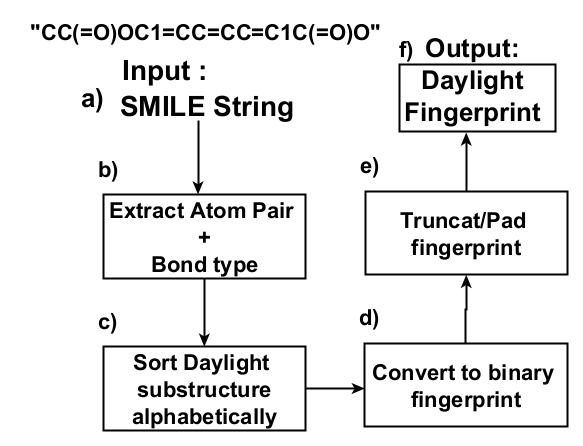}
  \caption{Daylight Fingerprint.}
  \label{Fig_Daylight_Fingerprint}
  \end{subfigure}%
  \begin{subfigure}{.33\textwidth}
  \centering
  \includegraphics[scale=0.19]{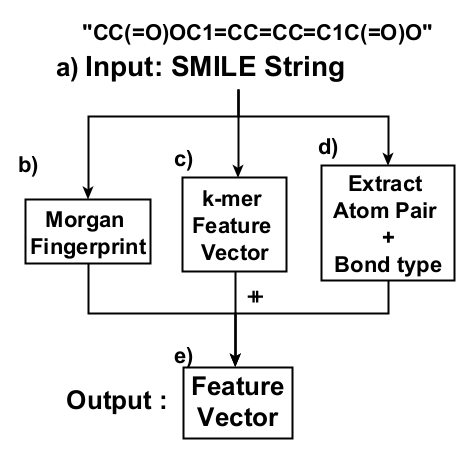}
  \caption{Proposed Method}
  \label{Fig_EMERGE}
  \end{subfigure}%
 \caption{Different methods for Feature Vector generation using SMILE String}
  \label{FV_Generation_Methods}
\end{figure}

The "Daylight Fingerprint" algorithm~\cite{james1995daylight}, as given in Algorithm~\ref{algo_daylight}, generates a binary fingerprint for a given SMILES string. It extracts atom pairs and bond types from the string, incrementing their counts in a dictionary. The counts are then converted to a binary representation, forming the fingerprint. The fingerprint is truncated or padded to the desired length. This unique binary representation captures the substructs present in the SMILES string. 
Figure~\ref{Fig_EMERGE} shows the process for generating the proposed Feature Vector.  
Figure~\ref{Fig_EMERGE} shows the process we use for generating the proposed Feature Vector which includes Morgan fingerprint with $k$-mer inclusion and Daylight fingerprint. 

\begin{algorithm}[h!]
\caption{Daylight Fingerprint}
\label{algo_daylight}
\scriptsize
\begin{algorithmic}[1]
\Function{GenerateDaylightFingerprint}{smiles, nBits=2048}
\State substructCnt $\gets$ {}
\For{$i \gets 0$ to len(smiles) - 2}
\State atom\_pair $\gets$ smiles[i:i+2]
\State bond\_type $\gets$ smiles[i+1:i+2]
\State substruct $\gets$ atom\_pair + bond\_type
\If{substruct $\in$ substructCnt}
\State substructCnt[substruct] $\mathrel{+}$= 1
\Else
\State substructCnt[substruct] $\gets$ 1
\EndIf
\EndFor
\State sortedSubstruct $\gets$ sort(substructCnt.keys())
\State fingerprint $\gets$ []
\For{substruct $\in$ sortedSubstruct}
\State substructBinary $\gets$ [int(bit) for bit in bin(substructCnt[substruct])[2:].zfill(32)]
\State fingerprint.extend(substructBinary)
\EndFor
\If{len(fingerprint) $\geq$ nBits}
\State fingerprint $\gets$ fingerprint[:nBits]
\Else
\State fingerprint $\gets$ fingerprint + [0] $\times$ (nBits - len(fingerprint))
\EndIf
\State \textbf{return} fingerprint
\EndFunction
\end{algorithmic}
\end{algorithm}

\section{Experimental Setup}~\label{sec_experimental_setup}
In this section, we report the dataset statistics. 
The detail regarding experimentation, including classifiers description along with evaluation metrics is reported in Section~\ref{sec_eval_metrics}. 
Moreover, the detail regarding the baseline models is also given in Section~\ref{sec_baselines}.
We obtained a dataset consisting of 6897 SMILES strings from the benchmark DrugBank dataset~\cite{shamay2018quantitative}. 
The objective is to classify drugs based on their subtypes, with a total of $188$ distinct subcategories being assigned as target labels.
The top 10 drug subcategories, obtained from the Food and Drug Administration (FDA) website~\footnote{\url{https://www.fda.gov/}}, are provided in Table~\ref{tbl_drug_subtypes}. To illustrate, Table~\ref{tbl_smiles_sample} presents an example of a SMILES string along with its corresponding attributes. We also performed t-SNE-based visualization of different embeddings as shown in Section~\ref{sec_viz}.

\begin{table}[h!]
    \centering
    \resizebox{0.82\textwidth}{!}{
    \begin{tabular}{ccccc}
    \toprule
    & & \multicolumn{3}{c}{String Length Statistics} \\
    \cmidrule{3-5}
    Drug Subcategory & Count & Min. & Max. & Avg. \\
    \midrule \midrule
        Others & 6299 & 2 & 569 & 55.4448 \\
        Barbiturate [EPC] & 54 & 16 & 136 & 51.2407 \\
        Amide Local Anesthetic [EPC] & 53 & 9 & 149 & 39.1886 \\
        Non-Standardized Plant Allergenic Extract [EPC] & 30 & 10 & 255 & 66.8965 \\
        Sulfonylurea [EPC] & 17 & 22 & 148 & 59.7647 \\
        Corticosteroid [EPC] & 16 & 57 & 123 & 95.4375 \\
        Nonsteroidal Anti-inflammatory Drug [EPC] & 15 & 29 & 169 & 53.6000 \\
        Nucleoside Metabolic Inhibitor [EPC]  & 11 & 16 & 145 & 59.9090 \\
        Nitroimidazole Antimicrobial [EPC] & 10 & 27 & 147 & 103.800 \\
        Muscle Relaxant [EPC] & 10 & 9 & 82 & 49.8000 \\
        \bottomrule
    \end{tabular}
    }
    \caption{Drug subtypes (Top $10$) extracted from FDA website. EPC $=>$ ``Established Pharmacologic Class".}
    \label{tbl_drug_subtypes}
\end{table}

\subsection{Baseline Models}
\label{sec_baselines}
In this section, we discuss various baseline techniques that were utilized to compare the outcomes with the proposed method.

\subsubsection{MACCS Fingerprint}
The binary fingerprint known as the MACCS fingerprint~\cite{keys2005mdl,durant2002reoptimization} makes use of predetermined substructs based on functional groups and ring systems typically present in organic compounds. The existence or absence of each substruct is encoded in the resulting binary vector.


\subsubsection{$k$-mers}
In the SMILES string, this approach uses a sequence-based embedding to express the frequencies of overlapping sub-sequences~\cite{kang2022surrogate} of length $k$. The SMILES string is broken up into overlapping sub-sequences of length $k$ using a sliding window, and the frequency of each sub-sequence is used to create an embedding. For our experiments, we use k=3.  
The frequency count for each $k$-mer is then taken to use for generating the feature vector.

\subsubsection{Weighted $k$-mers}
In order to improve the quality of the $k$-mers-based embedding, we adopt a weighted variant that uses Inverse Document Frequency (IDF) to give each $k$-mer in the embedding~\cite{ozturk2020exploring} a weight. Rare $k$-mers that exist in only a small number of SMILES strings are more informative than frequent $k$-mers that frequently appear in those strings. The frequency of each $k$-mer is therefore down-weighted using IDF based on the number of SMILES strings in which it appears. A weighted $k$-mers-based embedding that better reflects the distinctive characteristics of each SMILES string is the consequence of this. For our studies, $k=3$, and the Algorithm~\ref{algo_idf} provides the pseudocode for determining the weights for $k$-mers using IDF.





\begin{algorithm}[h!]
\begin{algorithmic}[1]
\scriptsize
\Function{WeightedKmers}{$kMersLst$}

    \State $totSamples \gets$ $\vert kMersLst \vert$ \Comment{$kMersLst:$ list of all $k$-mers}
    
    \State $weightsIDF \gets \{\}$ \Comment{Dictionary for set of $k$-mers}
    
    \For{$kmers$ \textbf{in} $kMersLst$}
        \For{$kVal$ \textbf{in} set($kmers$)}
            \If{$kVal$ not in $weightsIDF$}
                \State $weightsIDF[kVal] \gets 0$ \Comment{add new unique $k$-mers to dictionary}
            \EndIf
            
            \State $weightsIDF[kVal] ++ $ \Comment{increament corresponding $k$-mer count}
        \EndFor
    \EndFor
    \For{$kVal, ToT$ in $weightsIDF$}
        \State $weightsIDF[kVal] \gets$ log$(\frac{totSamples}{ToT})$ \Comment{log for \# of samples over $k$-mers count}
    \EndFor
    
    \Return $weightsIDF$
\EndFunction
\end{algorithmic}
\caption{Weighted $k$-mers Generation Using IDF}
\label{algo_idf}
\end{algorithm}

\subsection{Evaluation Metrics}
\label{sec_eval_metrics}
For our classification task, we employ a range of linear and non-linear classifiers, including SVM, Naive Bayes (NB), Multi-Layer Perceptron (MLP), K Nearest Neighbors (KNN), Random Forest (RF), Logistic Regression (LR), and Decision Tree (DT). Our evaluation metrics encompass average accuracy, precision, recall, weighted F1, macro F1, ROC-AUC, and classifier training runtime. To establish training and test sets, we randomly split our data with a $70-30\%$ distribution, and we conduct our experiments five times to obtain average outcomes. For hyperparameter tuning, we allocate $10\%$ of the training data as a validation set. To ensure reproducibility, we provide online access to our code and pre-processed dataset~\footnote{Available in the published version}.

\begin{table}[h!]
    \centering
    \resizebox{0.95\textwidth}{!}{
    \begin{tabular}{p{4.3cm}cp{3.5cm}c}
    \toprule
    SMILE String & Drug Name & Drug Subcategory & Solubility AlogPS \\
    \midrule
         \makecell[l]{ 
          [Ca++].CC([O-])=O.CC([O-])=O \\
         } 
          & Calcium Acetate & Non-Standardized Plant Allergenic Extract [EPC] & 147.0 g/l \\
 \bottomrule
    \end{tabular}
    }
    \caption{Randomly selected SMILES string example along with its drug name, drug subcategory, and Solubility AlogPS values.}
    \label{tbl_smiles_sample}
\end{table}

\subsection{Data Visualization}
\label{sec_viz}
\begin{figure}[h!]
  \centering
  \begin{subfigure}{.25\textwidth}
  \centering
  \includegraphics[scale=0.06]{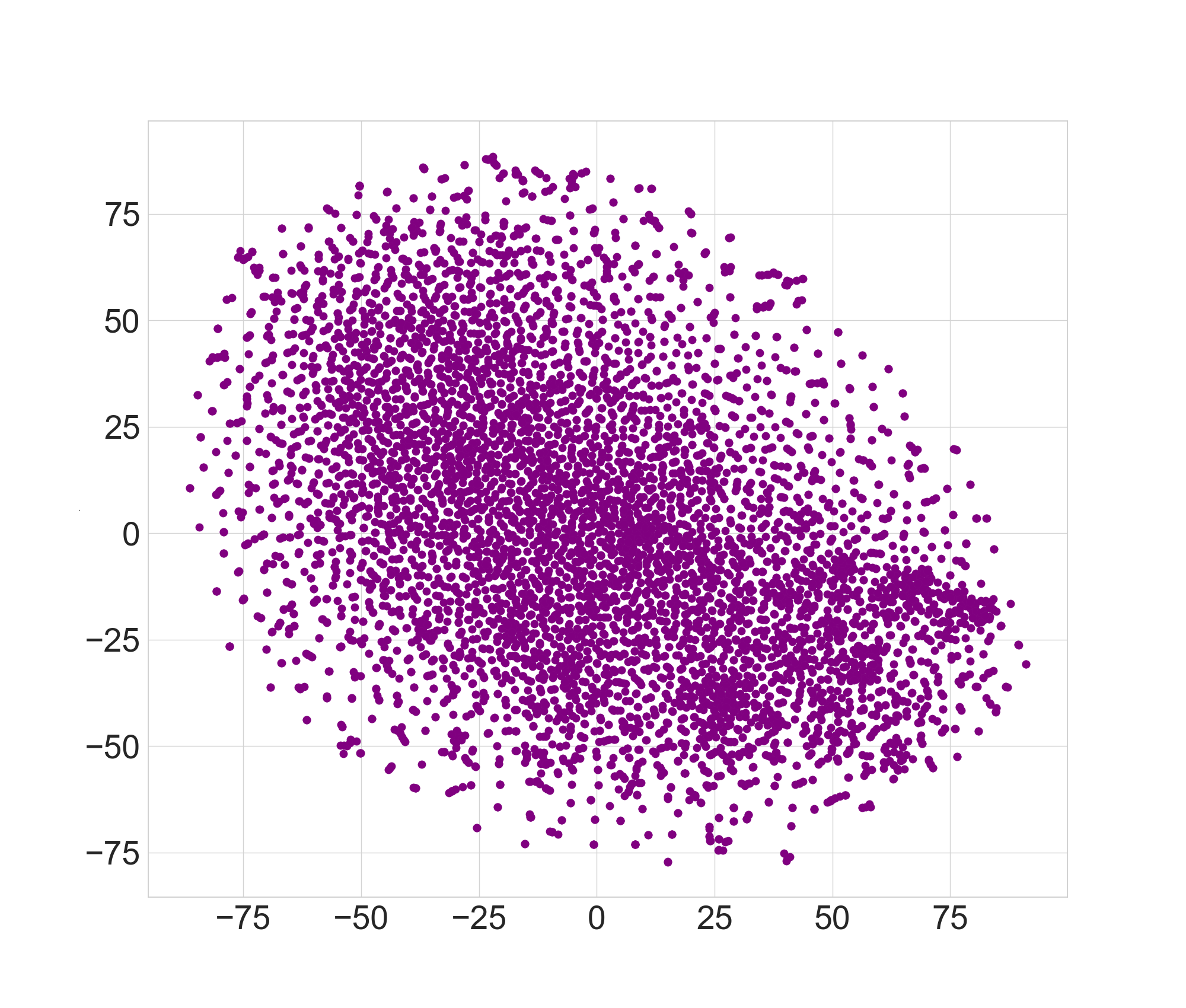}
  \caption{Morgan}
  \label{fig_tsne_kmer_cosine}
  \end{subfigure}%
  \begin{subfigure}{.25\textwidth}
  \centering
  \includegraphics[scale=0.06]{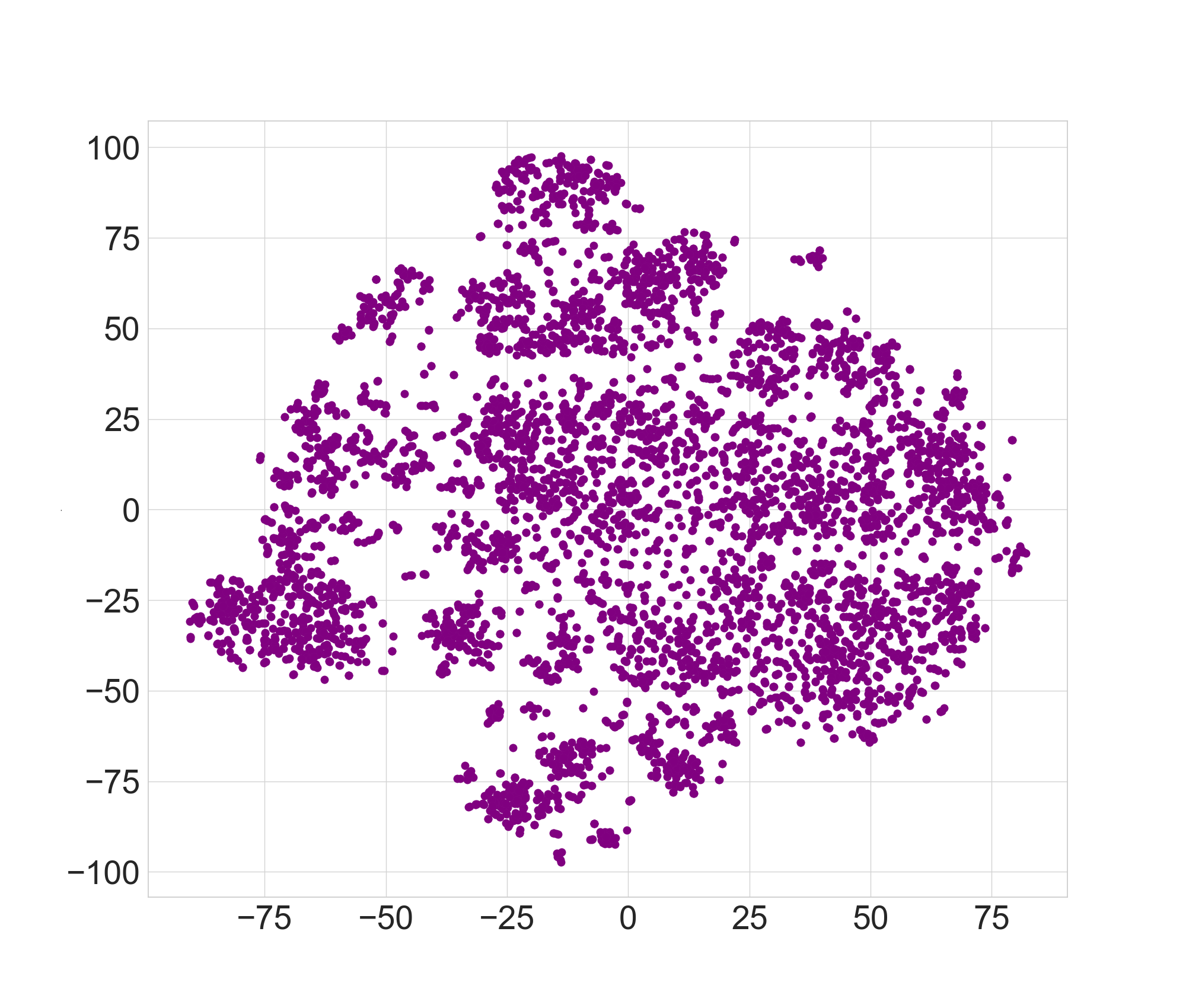}
  \caption{MACCS}
  \label{fig_tsne_kmer_laplacian}
  \end{subfigure}%
  \begin{subfigure}{.25\textwidth}
  \centering
  \includegraphics[scale=0.06]{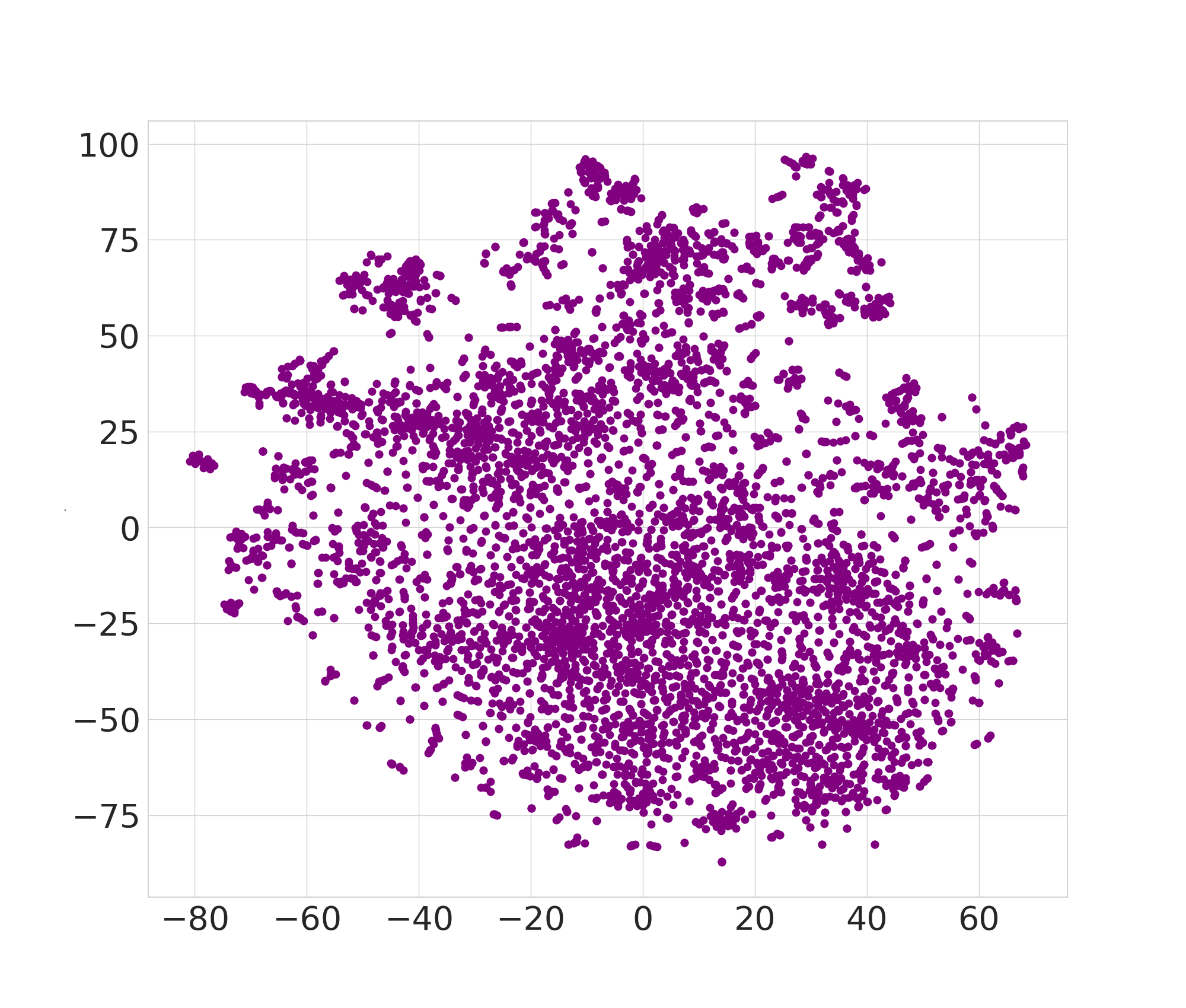}
  \caption{$k$-mers}
  \label{fig_tsne_kmer_gaussian}
  \end{subfigure}%
  \begin{subfigure}{.25\textwidth}
  \centering
  \includegraphics[scale=0.06]{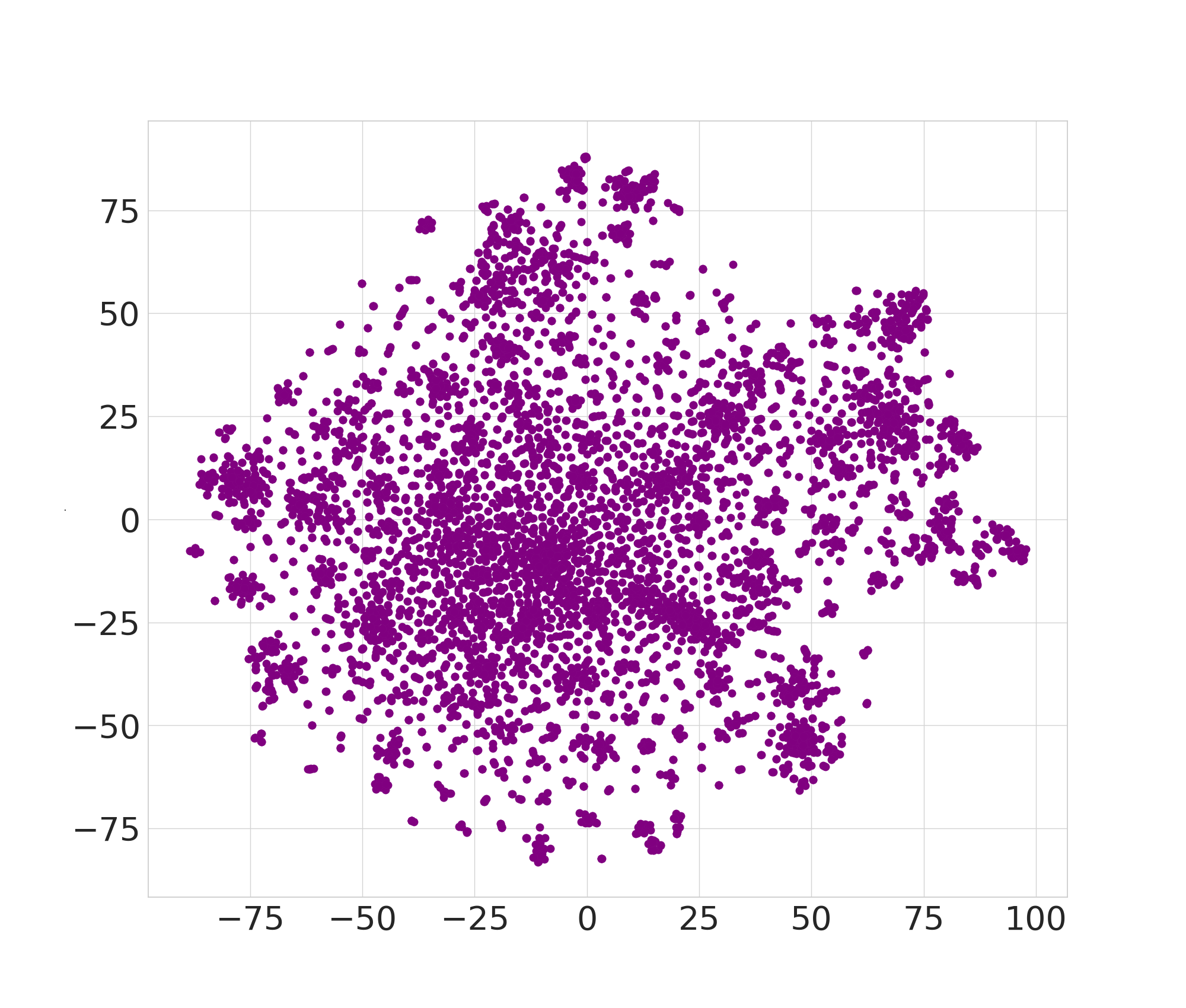}
  \caption{Weighted $k$-mers}
  \label{fig_tsne_kmer_gaussian}
  \end{subfigure}%
  \\
  \begin{subfigure}{.25\textwidth}
  \centering
  \includegraphics[scale=0.06]{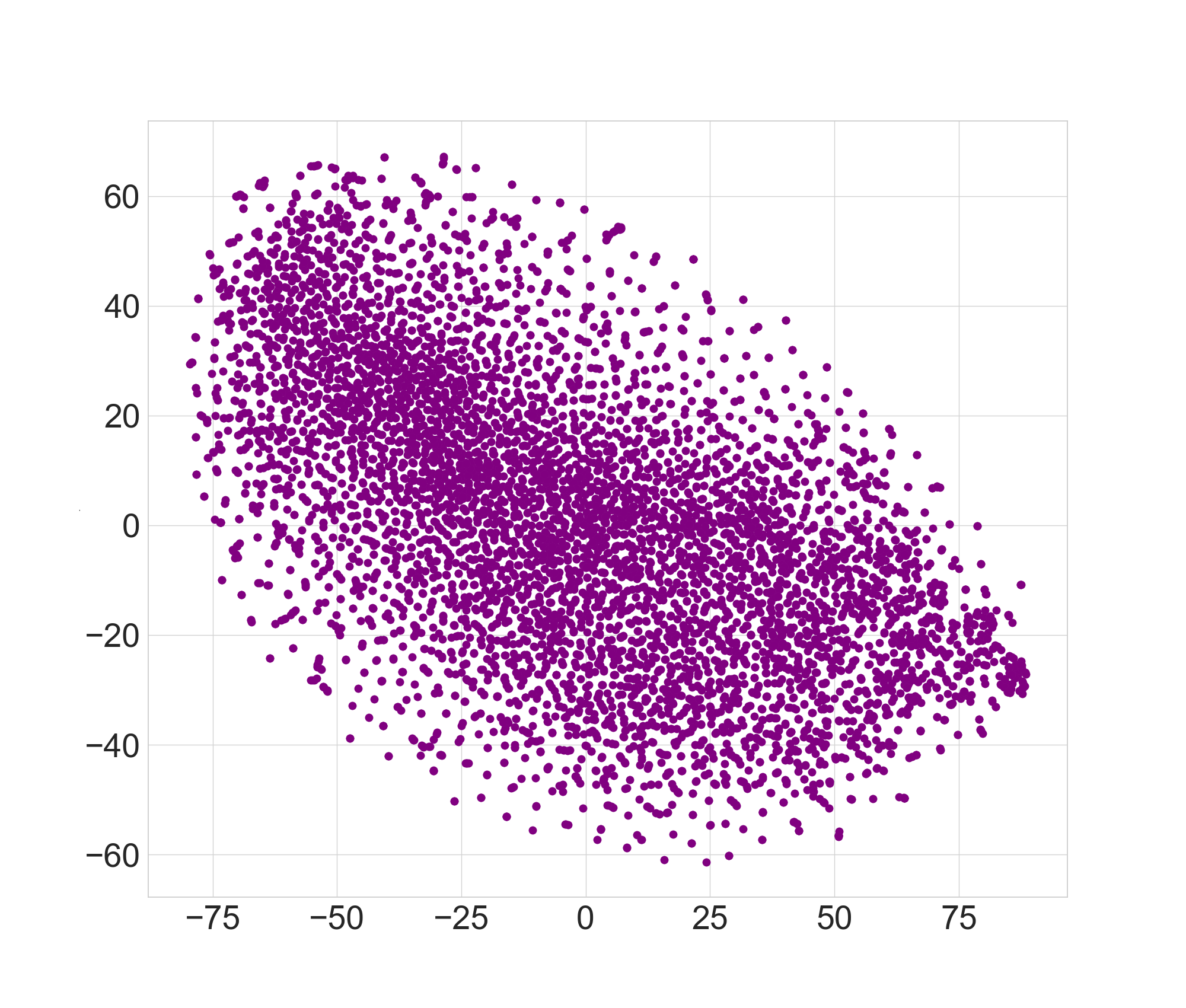}
  \caption{Daylight}
  \label{fig_tsne_kmer_gaussian}
  \end{subfigure}%
  \begin{subfigure}{.25\textwidth}
  \centering
  \includegraphics[scale=0.06]{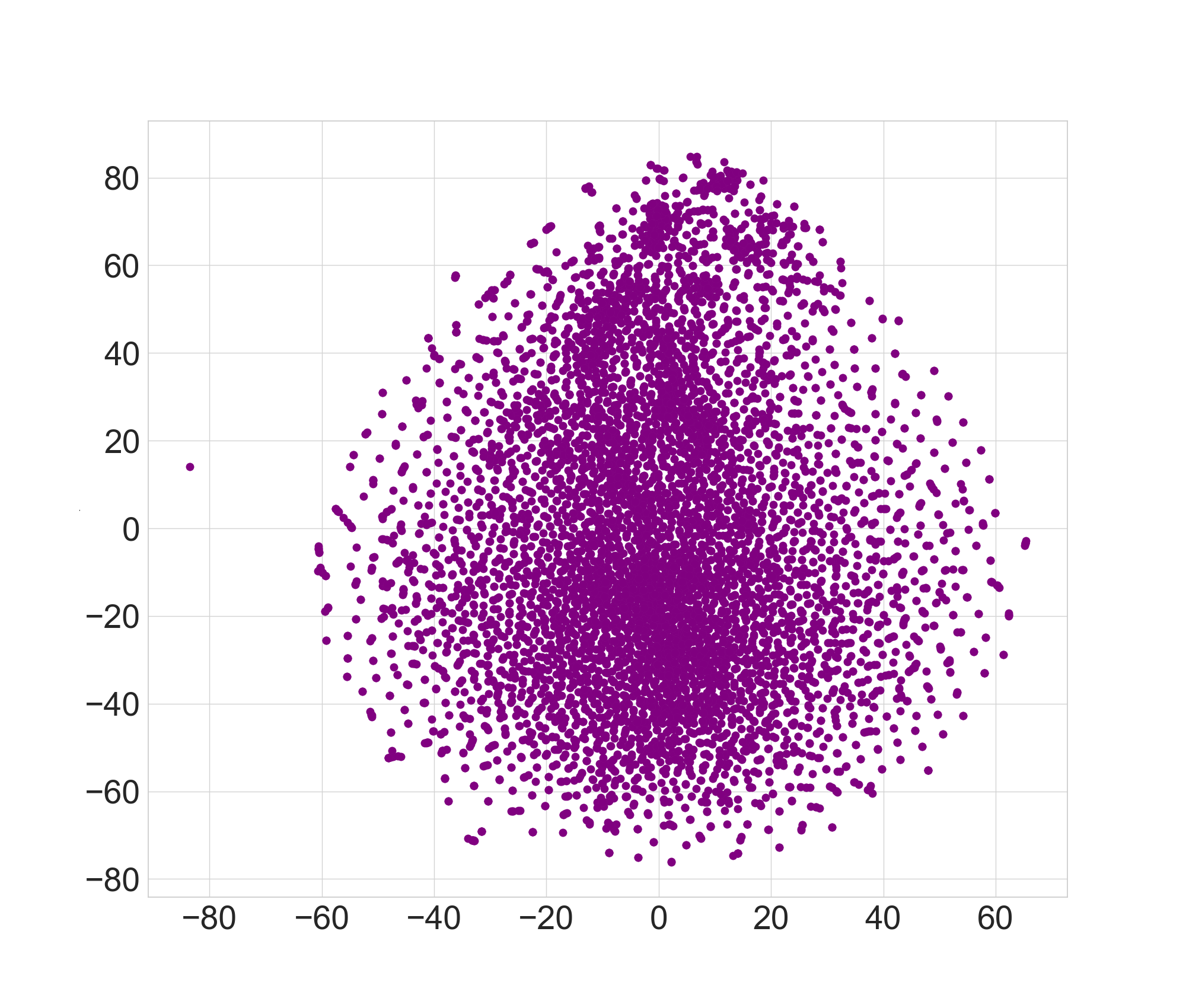}
  \caption{Morgan + $k$-mers}
  \label{fig_tsne_kmer_gaussian}
  \end{subfigure}%
  \begin{subfigure}{.25\textwidth}
  \centering
  \includegraphics[scale=0.06]{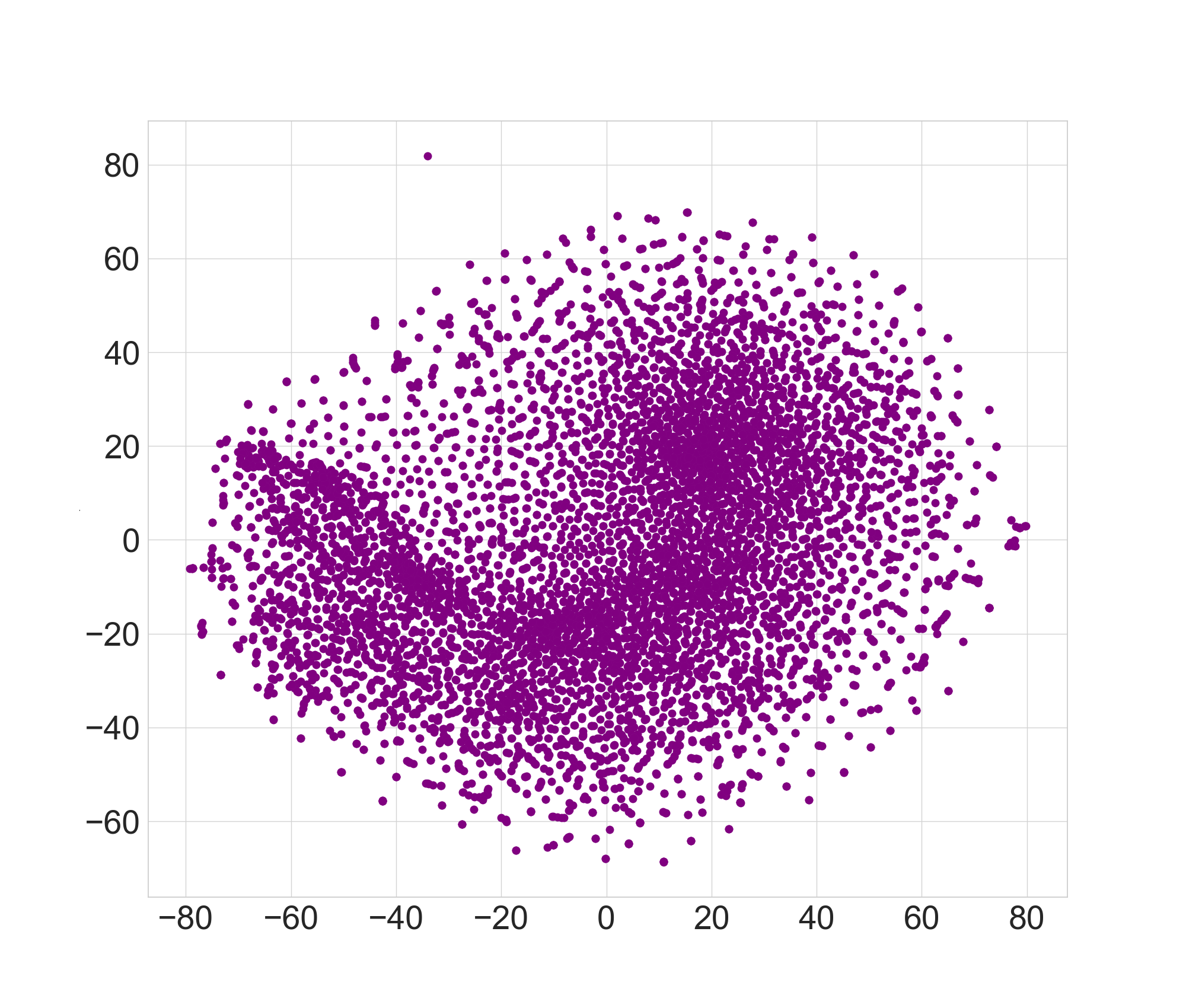}
  \caption{Morgan + DL}
  \label{fig_tsne_kmer_gaussian}
  \end{subfigure}%
  \begin{subfigure}{.25\textwidth}
  \centering
  \includegraphics[scale=0.06]{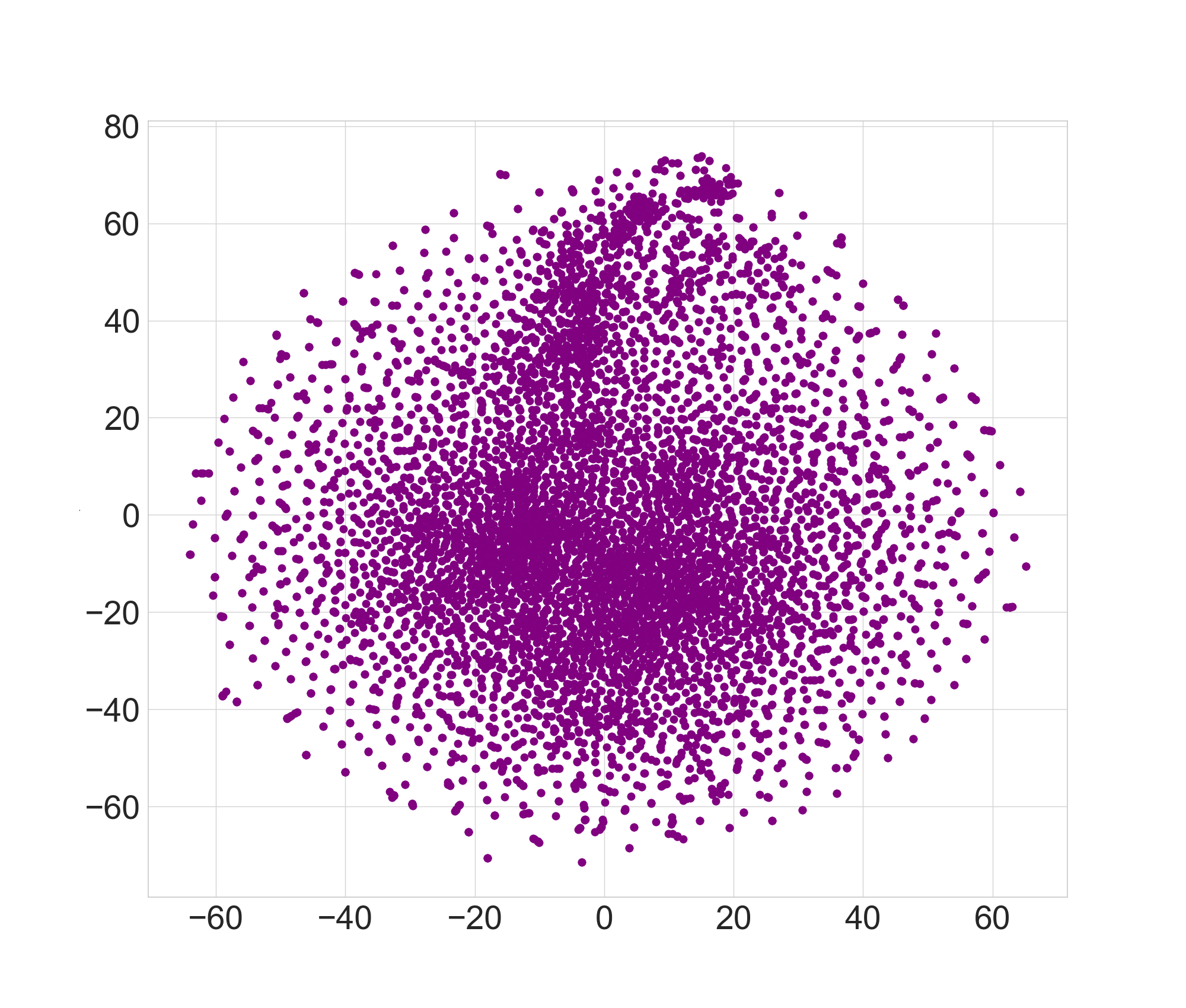}
  \caption{Morgan+$k$-mer+DL}
  \label{fig_tsne_kmer_gaussian}
  \end{subfigure}%
 \caption{The t-SNE plots for different feature embedding methods. The DL stands for Daylight.}
  \label{tsne_plots}
\end{figure}

We use the t-distributed Stochastic Neighbour Embedding (t-SNE) algorithm to create 2-dimensional representations of the different embeddings~\cite {van2008visualizing}. To have a visual inspection and determine whether different embedding strategies are keeping the structure of the data the t-SNE plots are generated. Figure~\ref{tsne_plots} shows the scatter plots produced by t-SNE for various embedding techniques. The MACCS fingerprint displays some clustering overall, which is similar for $k$-mers and weighted $k$-mer. On the other hand Morgan Fingerprint daylight are giving different scattered patterns. We can see the merged pattern with heavy inheritance from daylight when merged with Morgan. The proposed MERGE displays a mix of all in Figure~\ref{tsne_plots}(h), which is inherited clearly from Figure~\ref{tsne_plots}(f) and Figure~\ref{tsne_plots}(g).

\section{Results and Discussion}~\label{sec_results}
Table~\ref{tble_results_classification} presents the average classification results obtained from various methods and datasets, along with different evaluation metrics. 
We can observe that the proposed "Morgan + $k$-mers" method stands out with the highest accuracy (0.9162), precision (0.8541), recall (0.9162), and F1 score (0.8779). It also achieves a relatively low training time of 5.6350 seconds compared to other baselines. These results demonstrate its effectiveness in accurately classifying the datasets. Although the F1 (Macro) score is relatively low compared to some baselines, its overall performance is better, considering its high accuracy, precision, and recall.
Furthermore, the proposed method exhibits competitive performance in terms of ROC-AUC, indicating its ability to discriminate between positive and negative instances. Our Morgan Fingerprint + $k$-mers + Daylight Fingerprint performs the best in terms of ROC-AUC.

\begin{table}[h!]
    \centering
    \resizebox{0.85\textwidth}{!}{
    \begin{tabular}{@{\extracolsep{4pt}}cp{1cm}cccccp{1.3cm}p{2cm}}
    \toprule
        Embedding & Algo. & Acc. $\uparrow$ & Prec. $\uparrow$ & Recall $\uparrow$ & F1 (Weig.) $\uparrow$ & F1 (Macro) $\uparrow$ & ROC-AUC $\uparrow$ & Train Time (Sec.) $\downarrow$ \\
        \midrule \midrule                      
\multirow{7}{2.1cm}{MACCS Fingerprint~\cite{keys2005mdl,durant2002reoptimization}}                                
& SVM & 0.8705 & 0.8539 & 0.8705 & 0.8613 & 0.0520 & 0.5441 & 3.1812 \\
& NB & 0.2458 & 0.8473 & 0.2458 & 0.3698 & 0.0359 & 0.5224 & 0.5048 \\
& MLP & 0.8659 & 0.8444 & 0.8659 & 0.8547 & 0.0220 & 0.5175 & 21.0636 \\
& KNN & 0.9076 & 0.8447 & 0.9076 & 0.8741 & 0.0305 & 0.5107 & 0.0903 \\
& RF & 0.9057 & 0.8499 & 0.9057 & 0.8749 & 0.0344 & 0.5149 & 1.1254 \\
& LR & 0.9126 & 0.8331 & 0.9126 & 0.8710 & 0.0100 & 0.5000 & 3.2345 \\
& DT & 0.8227 & 0.8522 & 0.8227 & 0.8363 & 0.0457 & 0.5436 & \textbf{0.1100} \\

 \cmidrule{2-9}                                                                                     
 \multirow{7}{2.1cm}{$k$-mers~\cite{kang2022surrogate}}                           
 & SVM & 0.8190 & 0.8514 & 0.8190 & 0.8341 & 0.0413 & 0.5487 & 11640.03 \\
 & NB & 0.7325 & 0.8425 & 0.7325 & 0.7816 & 0.0247 & 0.5149 & 2348.88 \\
 & MLP & 0.8397 & 0.8465 & 0.8397 & 0.8426 & 0.0270 & 0.5311 & 7092.26 \\
 & KNN & 0.9101 & 0.8480 & 0.9101 & 0.8766 & 0.0429 & 0.5167 & 68.50 \\
 & RF & 0.9098 & 0.8449 & 0.9098 & 0.8740 & 0.0265 & 0.5075 & 655.47 \\
 & LR & 0.8885 & 0.8423 & 0.8885 & 0.8642 & 0.0461 & 0.5286 & 1995.11 \\
 & DT & 0.8429 & 0.8490 & 0.8429 & 0.8455 & 0.0397 & 0.5361 & 211.38 \\
 \cmidrule{2-9} 
\multirow{7}{2.1cm}{Weighted $k$-mers~\cite{ozturk2020exploring}}
 & SVM & 0.8219 & 0.8355 & 0.8219 & 0.8368 & 0.0451 & 0.5490 & 9926.76 \\
 & NB & 0.7490 & 0.8475 & 0.7490 & 0.7931 & 0.0360 & 0.5221 & 2564.96 \\
 & MLP & 0.8288 & 0.8511 & 0.8288 & 0.8392 & 0.0270 & 0.5345 & 7306.79 \\
 & KNN & 0.9122 & 0.8473 & 0.9122 & 0.8728 & 0.0307 & 0.5091 & 53.06 \\
 & RF & 0.9135 & 0.8455 & 0.9135 & 0.8758 & 0.0245 & 0.5067 & 619.65 \\
 & LR & 0.8928 & 0.8492 & 0.8928 & 0.8697 & \textbf{0.0595} & 0.5293 & 1788.37 \\
 & DT & 0.8420 & 0.8518 & 0.8420 & 0.8461 & 0.0445 & 0.5347 & 147.47 \\
 \cmidrule{2-9}   
\multirow{7}{2.1cm}{Daylight Fingerprint~\cite{james1995daylight}} 
 & SVM & 0.8562 & 0.8398 & 0.8562 & 0.8476 & 0.0165 & 0.5065 & 90.3683 \\
 & NB & 0.1591 & 0.8123 & 0.1591 & 0.2612 & 0.0058 & 0.5010 & 10.9286 \\
 & MLP & 0.8559 & 0.8371 & 0.8559 & 0.8462 & 0.0101 & 0.5041 & 53.3854 \\
 & KNN & 0.9115 & 0.8384 & 0.9115 & 0.8725 & 0.0120 & 0.5007 & 37.1265 \\
 & RF & 0.9112 & 0.8414 & 0.9112 & 0.8723 & 0.0138 & 0.5007 & 3.0294 \\
 & LR & 0.9129 & 0.8348 & 0.9129 & 0.8720 & 0.0134 & 0.5011 & 2.5398 \\
 & DT & 0.7958 & 0.8374 & 0.7958 & 0.8160 & 0.0111 & 0.5050 & 0.4753 \\

\cmidrule{2-9} 
\multirow{7}{2.1cm}{Morgan Fingerprint~\cite{james1995daylight}}
& SVM & 0.8564 & 0.8394 & 0.8564 & 0.8474 & 0.0245 & 0.5065 & 87.6153 \\
 & NB & 0.2792 & 0.8273 & 0.2792 & 0.4122 & 0.0089 & 0.5004 & 10.4096 \\
 & MLP & 0.8412 & 0.8373 & 0.8412 & 0.8391 & 0.0091 & 0.5065 & 42.6769 \\
 & KNN & 0.9094 & 0.8363 & 0.9094 & 0.8705 & 0.0120 & 0.5007 & 35.5932 \\
 & RF & 0.9105 & 0.8361 & 0.9105 & 0.8709 & 0.0131 & 0.5009 & 2.8328 \\
 & LR & 0.9117 & 0.8356 & 0.9117 & 0.8714 & 0.0159 & 0.5019 & 3.8399 \\
 & DT & 0.7934 & 0.8381 & 0.7934 & 0.8148 & 0.0163 & 0.5073 & 0.6120 \\
\cmidrule{2-9}   
\multirow{7}{2.1cm}{Morgan + $k$-mers (Ours)}
 & SVM & 0.8593 & 0.8459 & 0.8593 & 0.8522 & 0.0216 & 0.5088 & 97.1471 \\
 & NB & 0.4217 & 0.8413 & 0.4217 & 0.5573 & 0.0085 & 0.5011 & 10.3967 \\
 & MLP & 0.8249 & 0.8440 & 0.8249 & 0.8342 & 0.0096 & 0.5086 & 41.2894 \\
 & KNN & 0.9156 & 0.8460 & 0.9156 & 0.8778 & 0.0167 & 0.5026 & 35.3930 \\
 & RF & 0.9150 & 0.8453 & 0.9150 & 0.8772 & 0.0152 & 0.5017 & 2.6111 \\
 & LR & \textbf{0.9162} & \textbf{0.8541} & \textbf{0.9162} & \textbf{0.8779} & 0.0135 & 0.5010 & 5.6350 \\
 & DT & 0.8086 & 0.8449 & 0.8086 & 0.8262 & 0.0126 & 0.5066 & 0.6623 \\

\cmidrule{2-9}   
\multirow{7}{2.1cm}{Morgan + Daylight Fingerprint (Ours)}
 & SVM & 0.8593 & 0.8425 & 0.8593 & 0.8504 & 0.0247 & 0.5106 & 97.5363 \\
 & NB & 0.3472 & 0.8356 & 0.3472 & 0.4848 & 0.0084 & 0.5086 & 11.6528 \\
 & MLP & 0.8233 & 0.8389 & 0.8233 & 0.8309 & 0.0095 & 0.5081 & 46.4655 \\
 & KNN & 0.9131 & 0.8421 & 0.9131 & 0.8747 & 0.0136 & 0.5009 & 37.6461 \\
 & RF & 0.9126 & 0.8420 & 0.9126 & 0.8743 & 0.0188 & 0.5027 & 2.6711 \\
 & LR & 0.9140 & 0.8412 & 0.9140 & 0.8749 & 0.0153 & 0.5015 & 5.8494 \\
 & DT & 0.7958 & 0.8407 & 0.7958 & 0.8173 & 0.0117 & 0.5059 & 0.7014 \\

\cmidrule{2-9}   
\multirow{7}{2.1cm}{Morgan + $k$-mers + Daylight Fingerprint (Ours)}
 & SVM & 0.8521 & 0.8326 & 0.8521 & 0.8416 & 0.0216 & 0.5048 & 98.6409 \\
 & NB & 0.4354 & 0.8304 & 0.4354 & 0.5650 & 0.0087 & 0.5027 & 10.3506 \\
 & MLP & 0.8150 & 0.8326 & 0.8150 & 0.8236 & 0.0117 & 0.5120 & 41.1420 \\
 & KNN & 0.9093 & 0.8342 & 0.9093 & 0.8687 & 0.0176 & 0.5032 & 36.9900 \\
 & RF & 0.9088 & 0.8353 & 0.9088 & 0.8680 & 0.0129 & 0.5007 & 2.7349 \\
 & LR & 0.9101 & 0.8363 & 0.9101 & 0.8692 & 0.0152 & 0.5017 & 5.8551 \\
 & DT & 0.8114 & 0.8353 & 0.8114 & 0.8229 & 0.0152 & \textbf{0.5492} & 0.6160 \\

         \bottomrule
         \end{tabular}
    }
    \caption{Average Classification results (of $5$ runs) for different methods and datasets using different evaluation metrics. The best values are shown in bold.}
    \label{tble_results_classification}
\end{table}

\subsection{Statistical Significance}
To address concerns regarding the statistical significance of our results, we employed the student t-test. We calculated $p$-values using the averages and standard deviations (SD) from five runs, where each run involved different random data splits. It is worth noting that the SD values for all metrics were very small, typically below 0.002. As a result, we found that the $p$-values were less than 0.05, indicating statistical significance. 

\section{Conclusion}~\label{sec_conclusion}


In conclusion, we have presented a method for chemical representation with $k$-mers and fragment-based fingerprints for molecular fingerprinting, which is a novel method for generating molecular embeddings from SMILES strings. By combining the strengths of substruct counting, $k$-mers, and Daylight-like fingerprints, our method offers a more informative representation of chemical structures. Our experimental evaluations demonstrate the superiority of the proposed method over traditional methods, such as Morgan fingerprinting alone, in various cheminformatics tasks, including drug classification and solubility prediction. The integration of $k$-mers and Daylight-like fingerprints improves supervised analysis, making our method promising for molecular design and drug discovery. It advances the field of cheminformatics, offering new possibilities for molecular structure analysis and design.

\bibliographystyle{splncs04}
\bibliography{21}

\end{document}